\documentclass{JINST}
\usepackage{subfig}

\title{Energy Resolution studies for NEXT}

\author{C. A. B. Oliveira$^a$\thanks{Corresponding
author.}~, M. Sorel$^b$, J. Martin-Albo$^b$, J. J. Gomez-Cadenas$^b$, A. L. Ferreira$^a$ and J. F. C. A. Veloso$^a$\\
\llap{$^a$}i3N, Physics Department, University of Aveiro\\
  Campus de Santiago, 3810-193 Aveiro, Portugal\\
  \llap{$^b$}Instituto de F\'{i}sica Corpuscular (IFIC), CSIC and Universidad de Valencia\\
  Edificio Institutos de Investigaci\'{o}n, Apartado de Correos 22085, E-46071 Valencia - Spain\\
    E-mail: \email{carlos.oliveira@ua.pt} \\}

\abstract{This work aims to present the current state of simulations of electroluminescence (EL) produced in gas-based detectors with special interest for NEXT --- Neutrino Experiment with a Xenon TPC. NEXT is a neutrinoless double beta decay experiment, thus needs outstanding energy resolution which can be achieved by using electroluminescence. The process of light production is reviewed and properties such as EL yield and associated fluctuations, excitation and electroluminescence efficiencies, and energy resolution, are calculated. An EL production region with a 5~mm width gap between two infinite parallel planes is considered, where a uniform electric field is produced. The pressure and temperature considered are $10\textrm{ bar}$ and $293\textrm{ K}$, respectively. The results show that, even for low values of VUV photon detection efficiency, good energy resolution can be achieved: below 0.4\textrm{ }\% (FWHM) at $Q_{\beta\beta}=2.458\textrm{ MeV}$.}


\begin{document}

\section{Introduction}
\begin{figure}
\centering
\includegraphics[width=.80\textwidth]{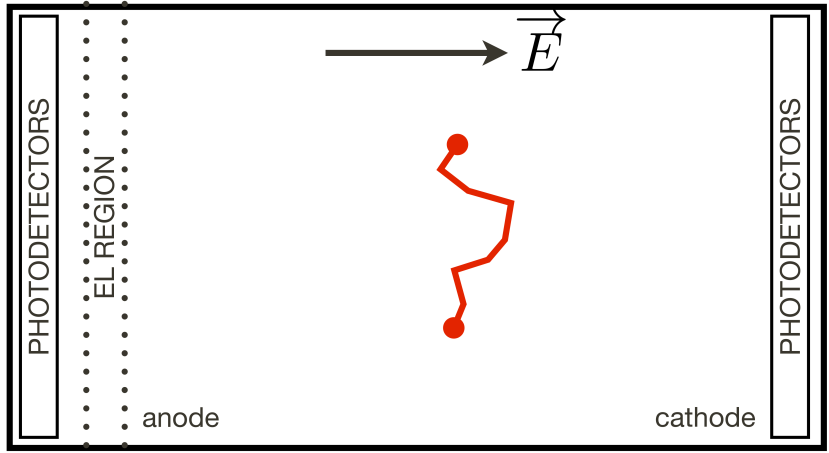}
\caption{Schematic representation of the Neutrino Experiment with a Xenon TPC detector and of the topological signature ``spaghetti with two meat balls'' of the two electrons emitted by the $\beta\beta^{0\nu}$ decay.}
\label{fig:schem}
\end{figure}

The Neutrino Experiment with a Xenon TPC (NEXT) will search for the $\beta\beta^{0\nu}$ decay in $^{136}$Xe using a 100 kg, high-pressure gaseous xenon, electrominescent time projection chamber (HPXe TPC). The project is approved for operation in the Canfranc Underground Laboratory (LSC), Spain. The TPC will have separated readout systems for calorimetry and tracking to facilitate both measurements (see Figure \ref{fig:schem}).

The two electrons emitted by the $\beta\beta^{0\nu}$ decay transfer their energy to the medium through ionization and excitation with the characteristic topological signature ``spaghetti with two meat balls''. The excitation energy results in the prompt emission of vacuum ultraviolet (VUV) light -- primary scintillation. The ionization tracks (positive ions and free electrons) left behind by the particles are prevented from recombination by a suitable drift electric field. Negative charge carriers -- primary electrons -- drift toward the TPC anode, entering a region with a more intense electric field created by two parallel meshes (EL region in Figure~\ref{fig:schem}). There, they are accelerated and collide with the gas atoms. Depending on the intensity of the electric field, electrons can either elastically collide with the atoms, or excite or even ionize them. As a result of the de-excitation of excited atoms, further VUV photons are generated isotropically -- secondary scintillation or electroluminescence (EL), Section \ref{teory}. Therefore, in such an electroluminescent TPC, both the scintillation and ionization processes ultimately produce VUV photons, to be detected with a photosensors array (photomultiplier tubes, PMTs) located behind the cathode. The detection of the primary scintillation light constitutes the start-of-event, $t_0$, whereas the detection of EL provides an energy measurement. Electroluminescent light is used also for tracking, as it is detected also by a second array of photosensors (e.g. multi-pixel photon counters, MPPCs) located behind and close to the EL region \cite{NEXTloi}.

Electroluminescence has already demonstrated in the past to be a very good option in order to achieve excellent energy resolution \cite{Bolotnikov,JMFSantosPMTdistance}, which is of particular importance in neutrinoless double beta decay searches. As an important example, high resolution spectrometry of low-energy X-rays is possible with Gas Proportional Scintillation Counters \cite{CondeGPSC,JMFSantosPortable}.

Because of this, it is important to understand the process of secondary light emission (electroluminescence)  and try to assess its different properties, namely excitation and electroluminescence efficiencies, EL yield and corresponding fluctuations. With these parameters in hand, it is possible to predict the light gain and energy resolution that are achievable with a particular detector geometry.  This assessment can be done through detailed simulations as explained in reference \cite{OliveirauE}. In this work we present results obtained applying detailed Monte Carlo simulations to the NEXT experiment conditions.

\section{Electroluminescence process}
\label{teory}

At pressures above few hundred Torr \cite{Tanaka}, the main channel of de-population of the excited atoms is through the formation of excimers --- electronically excited molecular states. Excimers, $R_2^{**}$, are formed through three-body collisions between one excited atom, $R^*$, and two atoms in the ground state, $R$:
\begin{equation}
R^*+2R \rightarrow R_2^{**} + R
\label{eq:excimerformation}
\end{equation}

The excimers are formed through process (\ref{eq:excimerformation}) in high vibrational states and, at the pressures studied in this work, collide with ground atoms, $R$, losing vibrational energy:
\begin{equation}
R_2^{**}+R \rightarrow R_2^{*} + R
\label{eq:vibdecay}
\end{equation}

The resultant excimers, $ R_2^*$, in a low vibrational state, emit VUV photons:
\begin{equation}
R_2^{*} \rightarrow  2R + h\nu
\label{eq:excimerdecay2}
\end{equation}

This results in the so-called ``second continuum'' centered at $173\textrm{ nm}$ and with a Full-Width-at-Half-Maximum (FWHM) of $14\textrm{ nm}$ \cite{Suzuki}. Further details about the electroluminescence process in pure noble gases can be found in Ref. \cite{OliveirauE}.

\section{Simulation}
\subsection{Toolkit}
The platform used to perform the simulations is the new C++ version of the microscopic technique of GARFIELD \cite{garfield}. This is a Monte Carlo technique which tracks the electrons at the atomic level using, currently, procedures and cross sections available in MAGBOLTZ 8.9.3 \cite{magboltz1,magboltz2}. 

The electron follows a vacuum trajectory between collisions, the path length being sampled according to an exponential distribution around the (energy-dependent) electron mean free path. The null-collision technique is used to consider the variations of the mean free path due to changes of the electron kinetic energy between collisions \cite{nullcoll}. In pure xenon, the program classifies the collisions as elastic, inelastic (electronic excitation) or ionization. It is possible to obtain information about each excited atom produced in the gas, namely: spatial position, time, excitation level and energy spent in the excitation.

\begin{figure}
\centering
\includegraphics[width=.80\textwidth]{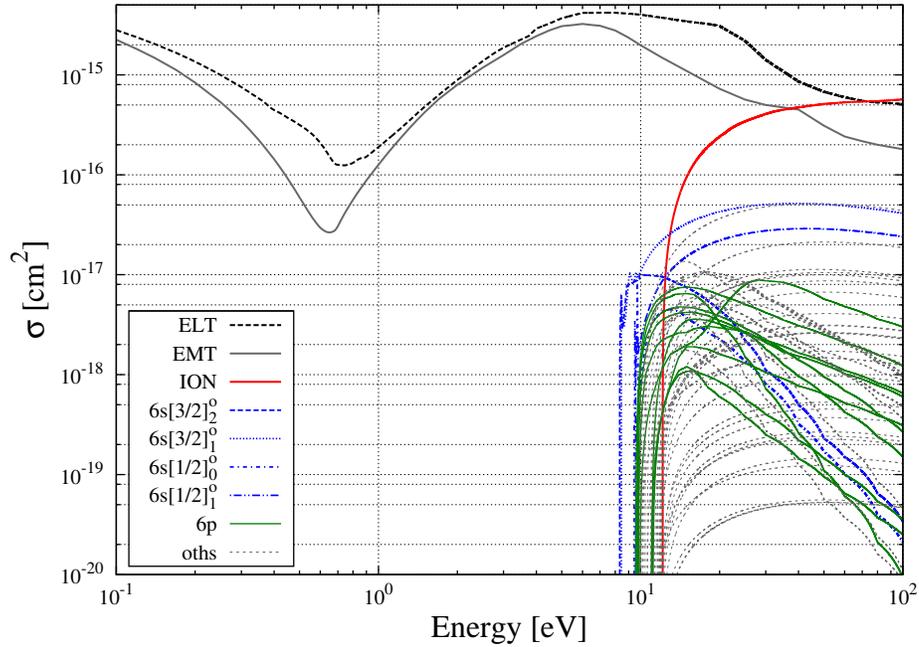}
\caption{Cross sections used by Magboltz 8.9.3 for Xe. The program includes detailed cross sections for elastic collisions, excitations (represented in 50 energy groups) and ionizations. Anisotropic elastic scattering is considered by using both the momentum transfer (EMT) and the total elastic (ELT) cross sections, shown as continuous and dashed gray lines, respectively. The blue lines refer to the first four excited states of Xe corresponding to the electronic configuration $5\mathrm{p}^56\mathrm{s}^1$ (Racah notation is used \cite{Racah}), the green lines correspond to the 10 levels of the $5\mathrm{p}^56\mathrm{p}^1$ configuration, and the gray dotted lines to other sparsely populated levels. The red line corresponds to ionization (ION).}
\label{fig:xsections}
\end{figure}

Figure \ref{fig:xsections} shows the cross sections used in our simulations. Detailed simulation of xenon excitations is possible by parametrizing the excited energy levels as a function of 50 energy groups.

An anisotropic angular distribution for elastic collisions is implemented by using both the momentum transfer (EMT) cross section and the total elastic (ELT) cross section according to the method described in Ref. \cite{anisotropic}.

\subsection{Model}
\label{sec:model}

\begin{figure}
\centering
\includegraphics[width=0.80\textwidth]{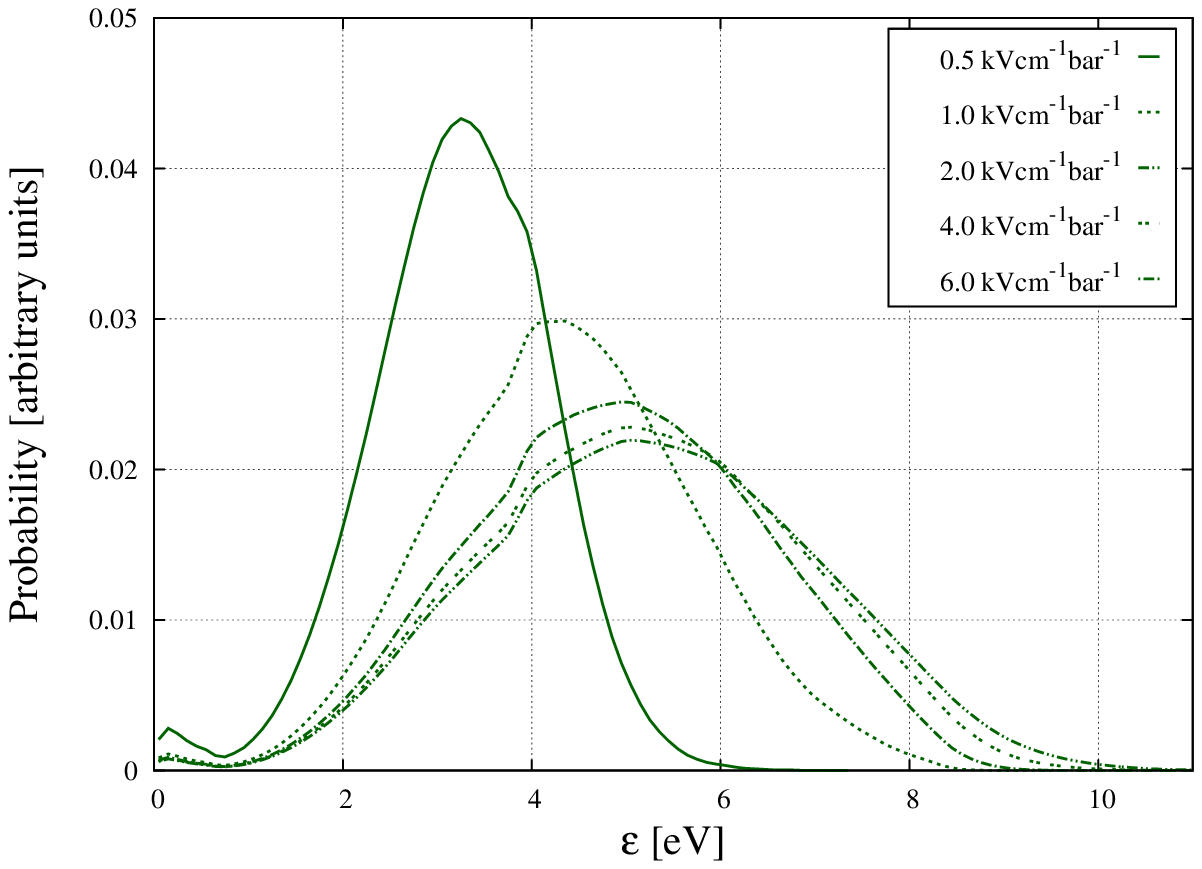}
\caption{Energy distributions of the electrons before each collision with the xenon atoms for different values of the reduced electric field. As the value of $\left(\frac{E}{p}\right)$ increases, the distribution gets broader and the peak shifts to higher energies.}
\label{fig:endist}
\end{figure}

Each primary electron is allowed to drift a distance of $d=5\textrm{ mm}$ under the influence of a uniform electric field. The effect of the field distortion near the meshes is not considered since it has no significant effect \cite{yop_argon}. We assume that the gas is at a pressure of $p=10\textrm{ bar}$ and at a temperature of $293\textrm{ K}$. A set of $N_\mathrm{e}=40,000$ primary electrons is used for each value of the voltage applied between the parallel planes, $V$. The starting direction of each primary electron is generated isotropically. The starting energy is generated according to the energy distribution calculated by Magboltz for the actual reduced electric field, $\left(\frac{E}{p}\right)=\frac{V}{dp}$, (see Figure \ref{fig:endist}).

We assume that every excited atom gives rise to the isotropic emission of a VUV photon. Its energy is generated according to a Gaussian distribution with the characteristics of the ``second continuum'' (see Section \ref{teory}).

The model was previously validated in Ref. \cite{OliveirauE} by comparing the simulations with available values of the EL yield, the excitation and the electroluminescence efficiencies. Earlier Monte Carlo results \cite{3d} and experimental measurements \cite{expy} in uniform electric field geometry are presented in Figure \ref{fig:yop_q} to illustrate the good agreement.

\section{Results}
\subsection{Efficiencies}
Excitation and electroluminescence efficiencies, $Q_\mathrm{exc}$ and $Q_\mathrm{EL}$, respectively, as a function of the reduced electric field are shown in Figure \ref{fig:yop_q}. The first quantity represents the fraction of energy that is supplied by the electric field to the primary electrons and that is spent in excitations \cite{3d}: 
\begin{equation}
Q_\mathrm{exc}=\frac{\displaystyle\sum\limits_{i=1}^{n_\mathrm{exc}}n^i\varepsilon^i_\mathrm{exc}}{edN_\mathrm{e}E}
\label{eq:qexc}
\end{equation}
In Eq.~(\ref{eq:qexc}), $n_\mathrm{exc}=50$ is the number of excitation groups available in Magboltz 8.9.3 for xenon, $n^i$ is the number of excitations of group $i$ produced by the set of $N_\mathrm{e}$ primary electrons, $\varepsilon_\mathrm{exc}^i$ is the energy of the $i^{th}$ excitation group, $E$ is the modulus of the applied electric field and $e$ is the charge of one electron. The second parameter, $Q_\mathrm{EL}$, is the ratio between the energy emitted in the form of VUV photons and the energy transferred from the field to the $N_\mathrm{e}$ electrons:
\begin{equation}
Q_\mathrm{EL}=\frac{\displaystyle\sum\limits_{i=1}^{n_\mathrm{exc}}\sum\limits_{j=1}^{n^i}\varepsilon^{i,j}_\mathrm{EL}}{edN_\mathrm{e}E}
\label{eq:qel}
\end{equation}
In Eq.~(\ref{eq:qel}), $\varepsilon^{i,j}_\mathrm{EL}$ is the energy of the VUV photon emitted after the $j^{th}$ excited atom of the $i^{th}$ group is produced. The VUV photon energy is randomly generated according to Section \ref{teory} parameters.

We note that $Q_\mathrm{EL}$ is always lower than $Q_\mathrm{exc}$. This is due mainly to the loss of vibrational energy by excimers before they emit a VUV photon (as explained in Section \ref{teory}). The fast increase of both efficiencies for $\left(\frac{E}{p}\right) >  5.0\textrm{ kVcm}^{-1}\textrm{bar}^{-1}$ is due to the contribution of secondary charges that start to be produced at high reduced field values.

\subsection{Electroluminescence yield}

\begin{figure}
\centering
\includegraphics[width=0.80\textwidth]{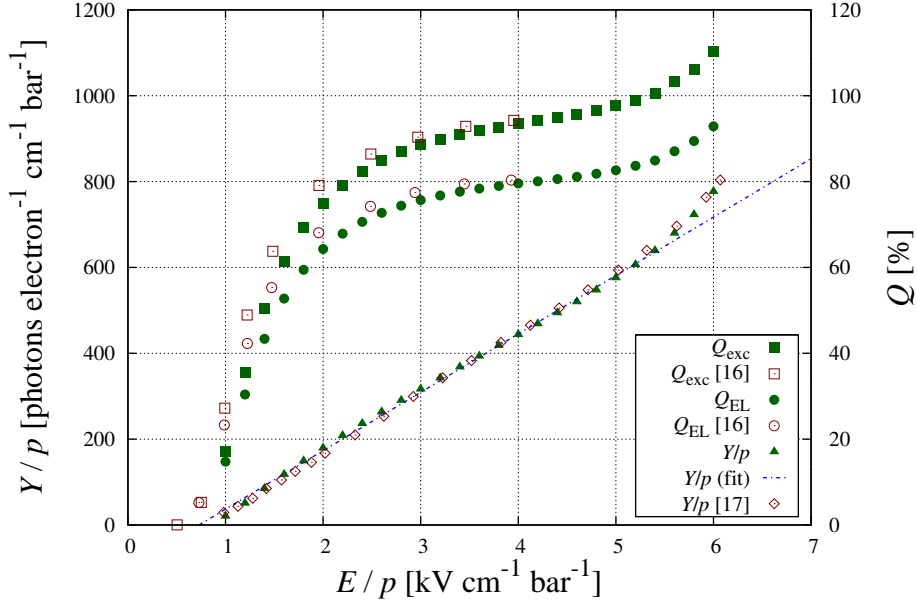}
\caption{Reduced EL yield, $Y/p$, as a function of the reduced electric field, $E/p$. Excitation efficiency, $Q_\mathrm{exc}$, and EL efficiency, $Q_\mathrm{EL}$, are also shown. Full symbols are the results of this work. Earlier Monte Carlo results for $Q_\mathrm{exc}$ and $Q_\mathrm{EL}$ \cite{3d}, as well as experimental measurements of the reduced EL yield \cite{expy}, are also shown (open symbols) for comparison.}
\label{fig:yop_q}
\end{figure}

\begin{figure}
\centering
\includegraphics[width=0.80\textwidth]{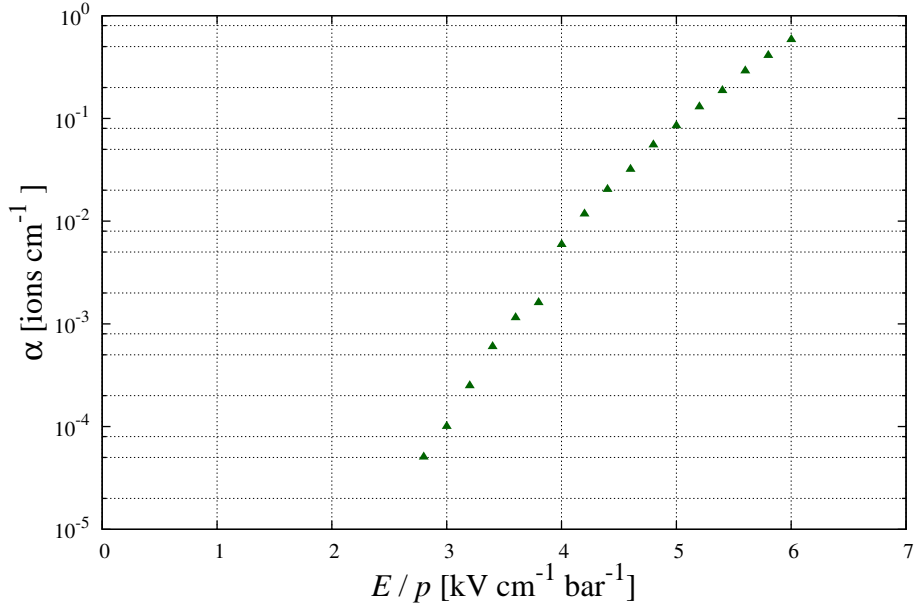}
\caption{Simulated first Townsend coefficient, $\alpha$ , as a function of the reduced electric field.}
\label{fig:alpha}
\end{figure}

The reduced electroluminescence yield, $\left(\frac{Y}{p}\right)$, as a function of the reduced electric field is also shown in Figure \ref{fig:yop_q}. The reduced electroluminescence yield is defined as the average number of VUV photons emitted per primary electron and per unit of drift length divided by the pressure of the gas, $p$. The behaviour of $\left(\frac{Y}{p}\right)$ with $\left(\frac{E}{p}\right)$ is approximately linear even if some ionization occurs. This happens for $\alpha<0.1\textrm{ ions}\cdot\textrm{cm}^{-1}$, being $\alpha$ the First Townsend coefficient (see Figure \ref{fig:alpha}), defined as the number of secondary charges produced when one electron crosses a distance of 1 cm. 

Performing a linear fit to the simulation results, we obtain the following dependence:
\begin{equation}
\left(\frac{Y}{p}\right)=\left(136\pm1\right)\left(\frac{E}{p}\right)-\left(99\pm4\right) \left[\textrm{photons electron}^{-1}\textrm{ cm}^{-1}\textrm{ bar}^{-1}\right]
\label{eq:yopfit}
\end{equation}
where $E/p$ is expressed in $\textrm{ kVcm}^{-1}\textrm{bar}^{-1}$. Eq.~(\ref{eq:yopfit}) is in good agreement with experimental data measured at $1\textrm{ bar}$ \cite{expy}:
\begin{equation}
\left(\frac{Y}{p}\right)=140\left(\frac{E}{p}\right)-116 \left[\textrm{photons electron}^{-1}\textrm{ cm}^{-1}\textrm{ bar}^{-1}\right]
\label{eq:yopexp}
\end{equation}

\subsection{Fluctuations and energy resolution}

The FWHM energy resolution, $R_\mathrm{E}$, of an EL detector (such as NEXT) corresponds to the quadrature sum of the contributions of the different processes happening in the detector:
\begin{equation}
\displaystyle R_\mathrm{E}=2\sqrt{2\ln{2}}\sqrt{\frac{\sigma_\mathrm{e}^2}{\overline{N}_\mathrm{e}^2}+\frac{1}{\overline{N}_\mathrm{e}}\left(\frac{\sigma_\mathrm{EL}^2}{\overline{N}_\mathrm{EL}^2}\right)+\frac{\sigma_\mathrm{ep}^2}{\overline{N}_\mathrm{ep}^2}+\frac{1}{\overline{N}_\mathrm{ep}}\left(\frac{\sigma_\mathrm{q}}{\overline{G}_\mathrm{q}}\right)^2}
\label{eq:re1}
\end{equation}

In Eq. (\ref{eq:re1}) the factor $2\sqrt{2\ln{2}}$ corresponds to the relation between the FWHM and the standard deviation, $\sigma$, of a given probability distribution ($\mathrm{FWHM}=2\sqrt{2\ln{2}}\sigma\simeq 2.35\sigma$). 
The first term of the expression is related to fluctuations in the number of primary charges created per event (that is a $\beta\beta^{0\nu}$ decay in NEXT), $N_\mathrm{e}$, the second to fluctuations in the number of EL photons produced per primary electron, $N_\mathrm{EL}$, the third reflects the variations in the number of photoelectrons extracted to the PMT photocathode per decay, $N_\mathrm{ep}$, and the fourth the distribution in the the PMT's single electron pulse height, $G_\mathrm{q}$.

\begin{figure}
\centering
\includegraphics[width=0.8\textwidth]{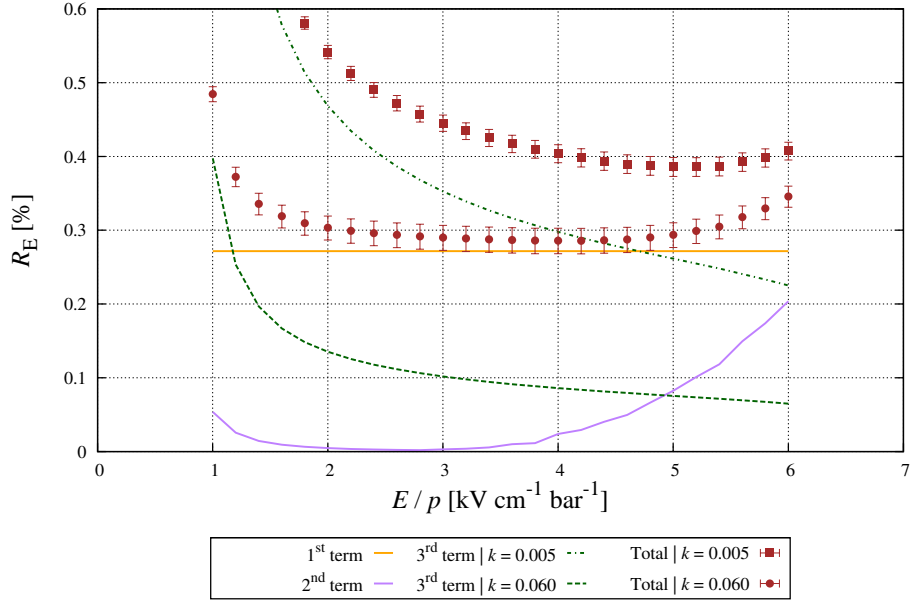}
\caption{FWHM energy resolution, $R_{E}$, as a function of the reduced electric field for two different scenarios: an optimistic (circles, $k=0.06$) and a conservative (squares, $k=0.005$). The values of $R_E$ if only each of the three terms of Eq. (\protect \ref{eq:re2}) would contribute are also shown.}
\label{fig:re}
\end{figure}

The primary charge fluctuations are well known and described by the Fano factor, $F=\sigma_\mathrm{e}^2/\overline{N}_\mathrm{e}$. In this work we consider $F=0.15$ \cite{nygren}. In literature are reported values between 0.13 and 0.17 for pure xenon \cite{fano1,fano2,fano3,fano4,w} which we used to estimate the overall energy resolution error bars presented in Figure~\ref{fig:re}.


The fluctuations associated with the electroluminescence production are described by the parameter $J$ defined as the relative variance in the number of emitted VUV photons per primary electron:
\begin{equation}
\displaystyle J=\frac{\sigma^2_{\mathrm{EL}}}{\overline{N}_\mathrm{EL}}
\label{eq:j}
\end{equation}

Concerning the third term, the conversion of VUV photons into photoelectrons follow a Poisson distribution and thus $\sigma_\mathrm{ep}^2=\bar{N}_\mathrm{ep}$. 

The fluctuations in the photoelectron multiplication gain within the PMT can typically be described by $\left(\frac{\sigma_\mathrm{q}}{\bar{G}_\mathrm{q}}\right)^2=1$ \cite{pmt_fluct}. Taking into account the previous relations, Eq. (\ref{eq:re1}) can be rewritten as follows \cite{JMFSantosPortable}:
\begin{equation}
\displaystyle R_\mathrm{E}=2.35\sqrt{\frac{F}{\overline{N}_\mathrm{e}}+\frac{1}{\overline{N}_\mathrm{e}}\left(\frac{J}{\overline{N}_\mathrm{EL}}\right)+\frac{2}{\overline{N}_\mathrm{ep}}}
\label{eq:re2}
\end{equation}

\begin{figure}
\centering
\includegraphics[width=0.8\textwidth]{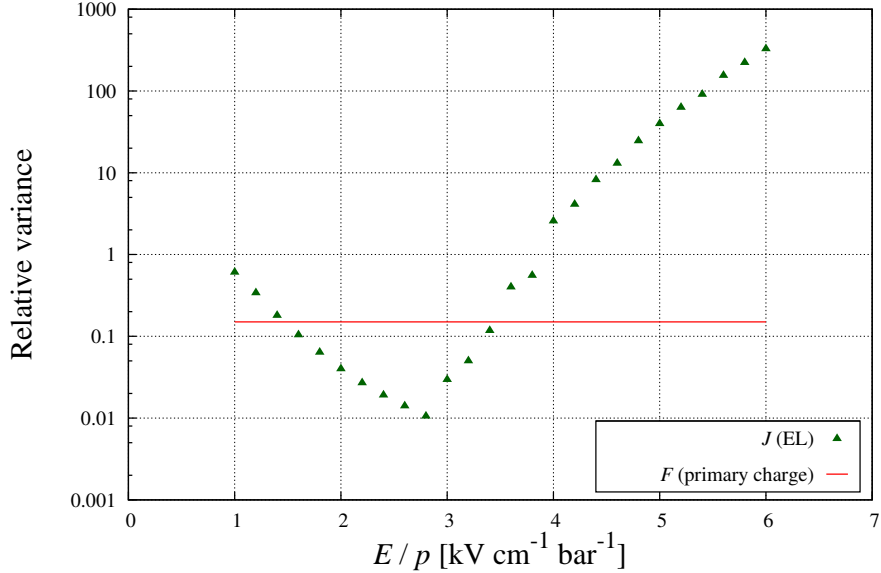}
\caption{Relative variance in the number of emitted EL photons, $J$, as a function of the reduced electric field. It is also shown the Fano factor -- the relative variance in the number of primary charges produced per event -- for high pressure xenon and a total deposited energy of $Q_{\beta\beta}=2.458\textrm{ MeV}$.}
\label{fig:j}
\end{figure}

\begin{figure}
  \centering
  \subfloat[$\left(\frac{E}{p}\right)=2\textrm{ kV cm}^{-1}\textrm{ bar}^{-1}$]{\label{fig:hist2}\includegraphics[height=0.4\textwidth]{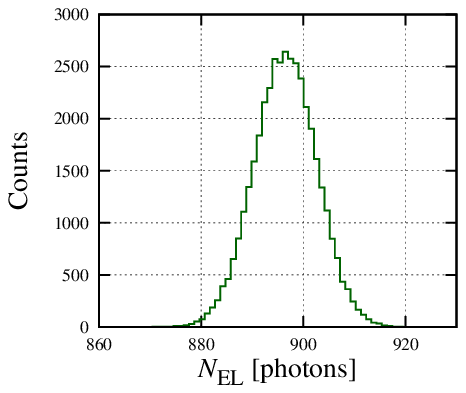}}\hspace{0.5cm}
  \subfloat[$\left(\frac{E}{p}\right)=3\textrm{ kV cm}^{-1}\textrm{ bar}^{-1}$]{\label{fig:hist3}\includegraphics[height=0.4\textwidth]{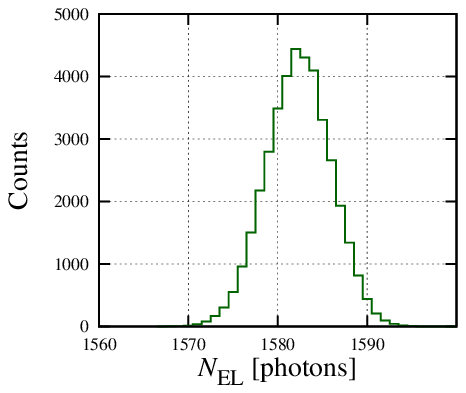}}\\
  \subfloat[$\left(\frac{E}{p}\right)=4\textrm{ kV cm}^{-1}\textrm{ bar}^{-1}$]{\label{fig:hist4}\includegraphics[height=0.4\textwidth]{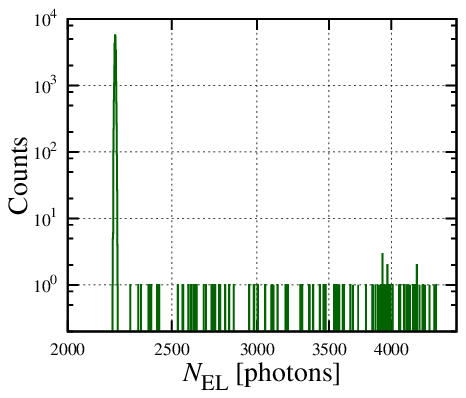}}\hspace{0.5cm}
  \subfloat[$\left(\frac{E}{p}\right)=5\textrm{ kV cm}^{-1}\textrm{ bar}^{-1}$]{\label{fig:hist5}\includegraphics[height=0.4\textwidth]{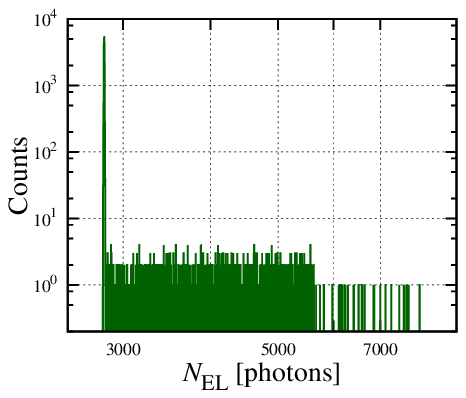}}
  \caption{Distribution in the number of emitted EL photons per primary electron, $N_\mathrm{EL}$, during the drift along $d= 5\textrm{ mm}$ of xenon at $10\textrm{ bar}$ and $293\textrm{ K}$, for different values of the reduced electric field. }
  \label{fig:hists}
\end{figure}

Usually $J$ is much smaller than $F$ and the energy resolution of a EL detector is attributed only to the fluctuations in the primary charge production and in the photon detection system (a plane of PMT's in the case of NEXT) -- first and third terms of Eq. (\ref{eq:re2}). Given that we deal with simulated data, in this work we can also calculate the parameter $J$ for the present geometry. Figure \ref{fig:j} shows $J$  as a function of the reduced electric field. As the electric field increases, the parameter $J$ decreases until secondary electrons begin to be produced and the avalanche fluctuations start to dominate. For $J$, values that are much smaller than the Fano factor, can be achieved in optimal conditions, namely for reduced electric fields between 1.5 and 3.5 $\textrm{kV cm}^{-1}\textrm{ bar}^{-1}$. In order to better understand the effect of avalanche fluctuations in EL fluctuations, in Figure \ref{fig:hists} we show the $N_\mathrm{EL}$ distributions for different values of the reduced electric field. The effect of secondary charges is clearly visible, since a long tail in the distribution appears for higher values of the electric field (Figures \ref{fig:hist4} and \ref{fig:hist5}). 

Using the data obtained by the simulation, we can estimate the energy resolution which can be achieved by the NEXT detector. We assume that all the energy of the two $\beta\beta^{0\nu}$ decay electrons, $Q_{\beta\beta}=2.458\textrm{ MeV}$, is deposited in the gas medium, with an average energy lost per ion pair formed of $w=21.9\textrm{ eV}$ \cite{nygren,w}:
\begin{equation}
\overline{N}_\mathrm{e}=\frac{Q_{\beta\beta}}{w}=\frac{2.458\times10^6\textrm{ eV}}{21.9\textrm{ eV}}\cong 112,237 \textrm{ electrons}
\label{eq:N}
\end{equation}

Taking into account the intensity of the electric fields in the drift and electroluminescence regions that are planned to be used, we assume that all the primary electrons arrive to the EL region and cross the first mesh without either any recombination or attachment to electronegative contaminants in the drift region.

The average number of photoelectrons produced in the plane of PMTs per decay, $\overline{N}_\mathrm{ep}$, can be obtained as:
\begin{equation}
\overline{N}_\mathrm{ep}=k\overline{N}_\mathrm{e}\overline{N}_\mathrm{EL}
\label{eq:Ne}
\end{equation}
where $k$ is the fraction of EL photons produced per $\beta\beta^{0\nu}$ decay that gives rise to the production of a photoelectron.

The FWHM energy resolution, $R_\mathrm{E}$, is shown in Figure \ref{fig:re} as a function of the reduced electric field for two different scenarios: an optimistic ($k=0.06$ \cite{nygren}) and a conservative ($k=0.005$ \cite{NEXTloi}). The contributions of each of the three terms of Eq. (\ref{eq:re2}) are also shown. The optimal energy resolution is achieved for a value of $\left(\frac{E}{p}\right)$ that is higher than the one marking the beginning of ionization, shown in Figure~\ref{fig:j} as the reduced field value minimizing the value of $J$. The reason is that the still small additional fluctuations introduced by the production of secondary charge are compensated by the increase in the number of EL photons produced per primary electron due to the increase of the EL yield. Also, and as expected, the reduced electric field value which gives the best energy resolution increases as $k$ decreases, and the best value of $R_\mathrm{E}$ is worst for lower values of $k$. For the experimental conditions considered here, the best energy resolution is obtained for $\left(\frac{E}{p}\right)$ between $\sim4.0$ and $\sim5.0\textrm{ kVcm}^{-1}\textrm{bar}^{-1}$, that is the value usually attributed to the ionization threshold of Xe \cite{ionthres}. 

\section{Conclusions}
The C++ version of the microscopic technique of Garfield, allows to gather information on the excited atoms produced during the drift of electrons through a xenon gas detector. Such simulations rely on detailed cross sections for elastic collisions, excitations and ionization as modelled by Magboltz 8.9.3. By considering that every excited atom leads to the emission of a vacuum ultraviolet (VUV) photon, we are able to accurately simulate electroluminescent (EL) light production in realistic detector geometry setups. 

In this work we consider EL light production in the NEXT $\beta\beta^{0\nu}$ experiment. We evaluate the EL yield and the corresponding fluctuations produced in such a setup, corresponding to a EL region given by a $5\textrm{ mm}$ uniform electric field gap of xenon at $293\textrm{ K}$ temperature and $10\textrm{ bar}$ pressure. We obtain that a FWHM energy resolution of 0.4\% or better at the $Q$ value of the $\beta\beta^{0\nu}$ reaction, $Q_{\beta\beta}=2.458\textrm{ MeV}$, can in principle be obtained, even considering conservative assumptions regarding the VUV photon detection efficiency.

In conclusion, our simulations indicate that the 1\% FWHM energy resolution goal of the NEXT experiment can in principle be met and even be improved upon, as far as fluctuations in the primary ionization charge, in the electroluminescence, and in the VUV photodetection processes are concerned. This energy resolution goal is one of the main figures of merit for the NEXT experiment \cite{NEXTloi}.

\acknowledgments
We acknowledge the support of the NEXT Collaboration (and the CONSOLIDER-INGENIO 2010 grant CSD2008-0037 (CUP)). This work was partially supported by project CERN/FP/109283/2009.
C. A. B. Oliveira was supported by a doctoral grant from FCT (Lisbon) with reference SFRH/BD/36562/2007.

\end{document}